\documentclass[aps,prl,superscriptaddress,twocolumn]{revtex4-2}

\usepackage{comment}
\usepackage[utf8]{inputenc}
\usepackage{amssymb}
\usepackage{amsmath}
\usepackage{amsfonts}
\usepackage{bm}
\usepackage{mathtools}
\usepackage{bbold}
\usepackage{upgreek}
\usepackage{xfrac}
\usepackage{soul}
\usepackage{bm}
\usepackage{xcolor}

\usepackage[colorlinks=true,citecolor=blue]{hyperref}
\hypersetup{colorlinks=true,citecolor=blue,linkcolor=blue,urlcolor=blue}
\usepackage{url}

\usepackage{txfonts}
\DeclareSymbolFont{matha}{OML}{txmi}{m}{it}
\DeclareMathSymbol{\varv}{\mathord}{matha}{118}
\usepackage{bbm}

\begin{document}

\title{Quantum Chebyshev Transform: Mapping, Embedding, Learning and Sampling Distributions}

\author{Chelsea A. Williams}
\thanks{These two authors contributed equally.}
\affiliation{Department of Physics and Astronomy, University of Exeter, Stocker Road, Exeter EX4 4QL, United Kingdom}
\affiliation{PASQAL, 7 Rue Léonard de Vinci, 91300 Massy, France}

\author{Annie E. Paine}
\thanks{These two authors contributed equally.}
\affiliation{Department of Physics and Astronomy, University of Exeter, Stocker Road, Exeter EX4 4QL, United Kingdom}
\affiliation{PASQAL, 7 Rue Léonard de Vinci, 91300 Massy, France}

\author{Hsin-Yu Wu}
\affiliation{Department of Physics and Astronomy, University of Exeter, Stocker Road, Exeter EX4 4QL, United Kingdom}
\affiliation{PASQAL, 7 Rue Léonard de Vinci, 91300 Massy, France}

\author{Vincent E. Elfving}
\email{vincent.elfving@pasqal.com}
\affiliation{PASQAL, 7 Rue Léonard de Vinci, 91300 Massy, France}

\author{Oleksandr Kyriienko}
\email{o.kyriienko@exeter.ac.uk}
\affiliation{Department of Physics and Astronomy, University of Exeter, Stocker Road, Exeter EX4 4QL, United Kingdom}
\affiliation{PASQAL, 7 Rue Léonard de Vinci, 91300 Massy, France}

\date{\today}

\begin{abstract}
We develop a paradigm for building quantum models in the orthonormal space of Chebyshev polynomials. We show how to encode data into quantum states with amplitudes being Chebyshev polynomials with degree growing exponentially in the system size. Similar to the quantum Fourier transform which maps computational basis space into the phase (Fourier) basis, we describe the quantum circuit for the mapping between computational and Chebyshev spaces. We propose an embedding circuit for generating the orthonormal Chebyshev basis of exponential capacity, represented by a continuously-parameterized shallow isometry. This enables automatic quantum model differentiation, and opens a route to solving stochastic differential equations. We apply the developed paradigm to generative modeling from physically- and financially-motivated distributions, and use the quantum Chebyshev transform for efficient sampling of these distributions in extended computational basis.
\end{abstract}

\maketitle

\textit{Introduction.---}Quantum computing is moving from theory to practice. Currently, researchers are actively seeking quantum advantage, and various algorithms motivate this goal~\cite{NielsenChuang}. Among these algorithms are Shor's order-finding~\cite{shor1994algorithms}, linear system solvers (HHL)~\cite{harrow2009quantum}, Monte-Carlo pricing~\cite{Rebentrost2018}, and ground state preparation for chemistry and materials~\cite{McArdle2020}. All aforementioned protocols rely on subroutines that involve quantum Fourier transform (QFT) and quantum phase estimation~\cite{NielsenChuang,Aspuru-Guzik2005}. QFT is crucial for the selection of an appropriate basis in which quantum information is processed~\cite{Browne2007}. From an intuitive perspective, these protocols can be envisioned as operating in the phase (reciprocal) space and subsequently transformed into the computational (real) space to enable efficient readout. An analogous situation arises in condensed matter theory, where describing lattice systems in a suitable $k$-space basis can yield significant computational advantages and enable scalable modeling, circumventing the need for a large computational mesh. We posit that similar considerations apply when constructing quantum models and algorithms.

Quantum machine learning (QML) is a nascent field that emerged at the interface of artificial intelligence and quantum information processing~\cite{Benedetti2019,Biamonte2017}. It focuses on manipulating quantum data, either embedded from classical datasets or generated in quantum experiments~\cite{Cerezo2022rev}. Inspired by the success of classical machine learning in tasks like image recognition~\cite{Krizhevsky2017} and language processing~\cite{radford2018improving}, QML aims to build quantum models that efficiently solve these problems. In simple terms, each QML model is a quasi-probability distribution created by a quantum circuit (followed by optional measurement post-processing) \cite{Benedetti2019,SchuldSweke2021}. The goal is to prepare expressive and trainable quantum models using high-dimensional Hilbert spaces~\cite{schuld2021supervised}. However, this is challenging because high expressivity can lead to poor trainability \cite{ZoeHolmes2022}, manifested in barren plateaus \cite{McClean2018,Cerezo2021NatComm}, exponential cost concentration \cite{Arrasmith2022}, local minima trapping \cite{Anschuetz2022}, and modified bias-variance trade-off \cite{Banchi2021,Caro2021encodingdependent}. Robust approaches to handle this are currently under development. They include embedding symmetries \cite{Larocca2022PRXQ,Meyer2023PRXQ}, limiting entanglement \cite{Patti2021,Sack2022PRXQ}, enabling overparametrization \cite{Larocca2021overparametrization,Larocca2022diagnosingbarren}, adapting ansatze \cite{Grimsley2023}, crafting cost functions \cite{Cerezo2021NatComm}, dropout \cite{Kobayashi2022}, and including differential constraints via physics-informed learning \cite{kyriienko2021solving,Paine2021,heim2021quantum}. The crucial point for building a high-performing quantum model is choosing an appropriate basis, coming from a selected quantum feature map \cite{Caro2021encodingdependent}.

Unlike quantum computing protocols that aim to achieve speed up in terms of runtime and gatecount scaling, QML goes beyond operational advantages, and offers conceptually new benefits \cite{Schuld2022PRXQ}. While quantum-enabled processing bears its own fundamental importance, particular advantages of QML can come from improved learning~\cite{Caro2022,caro2022outofdistribution,bowles2023contextuality} and sampling~\cite{hangleiter2023computational,Arute2019,Zhong2020,Madsen2022}. The learning advantage originates from generalization on fewer data points~\cite{Caro2022}, achieved in convolutional networks applied to magnetic phase recognition and entanglement detection~\cite{Cong2019,Caro2022,LiuPollmann2023qcnn}. For example, this was recently successfully applied to learning from quantum data, executed on a 40-qubit device~\cite{Huang2022}. The sampling advantage comes from the inherently different procedure of generating a sample from quantum distributions as compared to classical procedure \cite{hangleiter2023computational}. While classical sampling requires integrating and inverting probability distributions (with complexity increasing for correlated stochastic variables \cite{Kyriienko2022}), quantum devices are sampled directly via projective measurements in the computational basis. This underpins the successful demonstration of quantum supremacy based on random circuits~\cite{Arute2019}, also repeated recently with 70-qubit random circuits~\cite{google2023supremacy}. While to date sampling advantage has not been realized for relevant problems, this approach holds promise for generative modeling tasks \cite{Benedetti2019,Liu2018,PerdomoOrtiz2018,Coyle2020,Zoufal2019,Romero2021,Paine2021,Kyriienko2022,GiliPRA2023,gili2023generative}. The crucial point for efficient quantum generative modeling is an ability to transform a quasi-probability distribution for learning-favorable basis into the computational basis with fast and easy sampling \cite{Mazzola2021}.
\begin{figure*}[t]
\begin{center}
\includegraphics[width=1.0\linewidth]{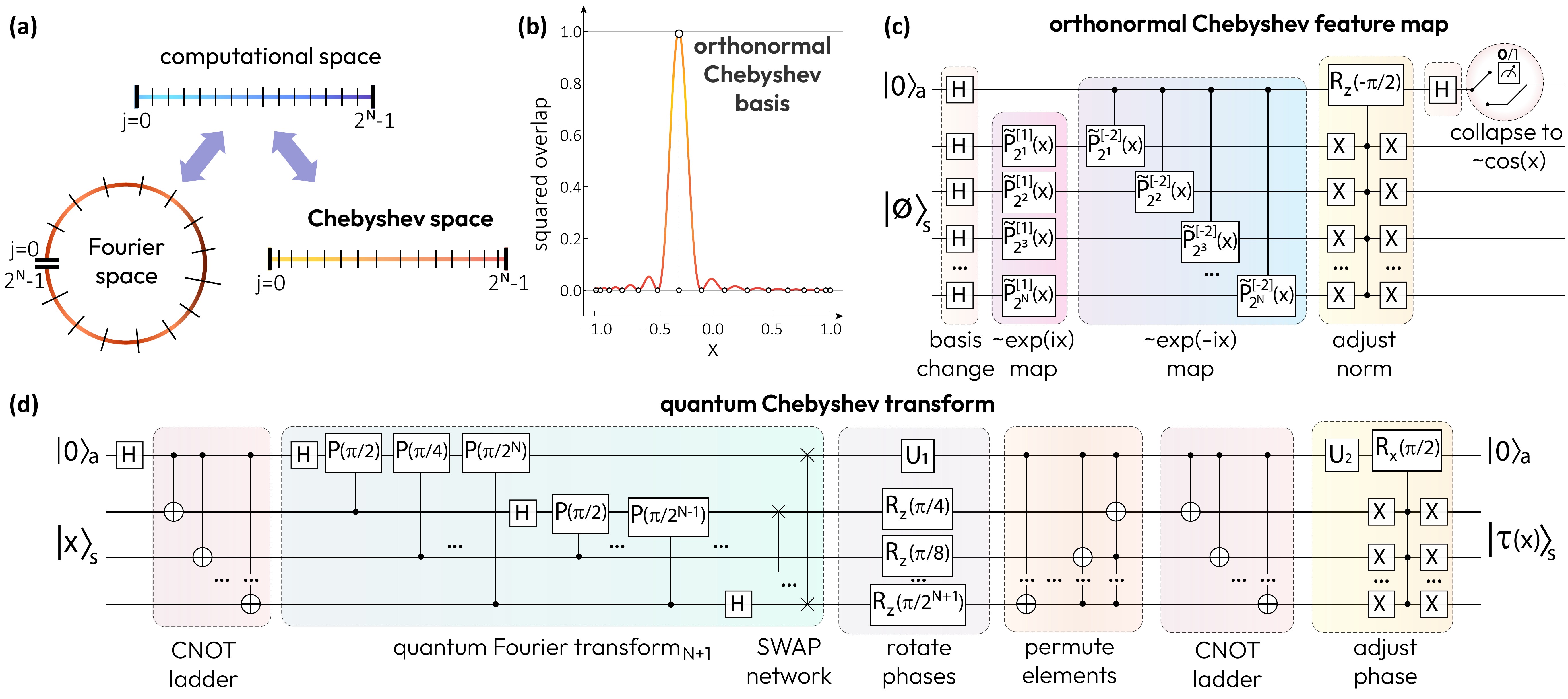}
\end{center}
\caption{\textbf{Workflow for quantum Chebyshev-based model building. (a)} Visualization for mapping bitstrings from the computational basis (labelled by integers $j \in [0,2^N - 1]$) into the Chebyshev space with non-equidistant nodes given by $x_{j}^{\mathrm{Ch}} = \cos\left(\pi(2j+1)/2^{N+1} \right)$, or the Fourier space with an equidistant grid of phases (sectors on a circular open-ended loop with $2\pi j/2^N$ arcs). \textbf{(b)} Squared overlap between two Chebyshev states $|\langle \tau(x')|\tau(x) \rangle|^2$, where we choose $x' = x_{7}^{\mathrm{Ch}}$, showing that the basis is orthonormal at the nodes. Notably, the Chebyshev overlaps are real, unlike the complex overlaps between Fourier states, but lead to similar mid-point behavior. \textbf{(c)} Quantum Chebyshev feature map that creates a Chebyshev state via $x$-parametrized isometry --- sequence of phase feature maps that embed complex exponents, controlled rotation to adjust the norm of the zero frequency term, and post-selecting the sum term with ancilla measured in $0$. Scaled single-qubit phase shift gates in the feature map circuit are defined as $\tilde{\mathrm{P}}^{[s]}_l(x)=\mathrm{diag}\{1,\exp(i s 2^N \arccos(x)/l)\}$, where $l$ grows exponentially as $2^j$, with $j$ being the qubit number, and $s$ takes values of $1$ and $-2$. Here, $\mathrm{H}$ and $\mathrm{X}$ are Hadamard and Pauli $\hat{X}$ gates. \textbf{(d)} Quantum Chebyshev transform ($\hat{\mathcal{U}}_{\mathrm{QChT}}$) circuit which maps computational basis states $\{ |x_j\rangle \}_{j=0}^{2^N -1}$ into Chebyshev states $\{ |\tau(x_j^{\mathrm{Ch}})\rangle \}_{j=0}^{2^N -1}$. The transform involves a QFT circuit applied to $N+1$ qubits ($N$-qubit system plus one ancillary qubit), followed by phase-adjusting and permutation circuits. Here, we use a standard phase shift gate defined as $\mathrm{P}(\phi)=\mathrm{diag}\{1,\exp(i\phi)\}$. We note that local phase rotations $\mathrm{U}_1 = \mathrm{P}(-\pi/2^{N+1}) \mathrm{R}_\mathrm{Z}(-\pi (2^{N}-1)/2^{N+1})$ and $\mathrm{U}_2 = \mathrm{P}(-\pi/2) \mathrm{R}_\mathrm{Y}(-\pi/2)$ acting on the ancilla can be combined into a single gate.
}
\label{fig:workflow}
\end{figure*}

In this paper, we propose conceptually distinct tools that enable quantum model building based on the Chebyshev basis \cite{edeclaration}. Chebyshev polynomials enable optimal function approximation with respect to $L_{\infty}$ norm, and often lead to practical advantages over the Fourier basis \cite{trefethen2019approximation} (which is suitable for harmonic models but performs poorly for polynomial decomposition). We propose an orthonormal Chebyshev map as a quantum circuit (formally, isometry \cite{RabanIten2016}) parameterized by continuous variable $x$, which creates quantum basis states with amplitudes corresponding to Chebyshev polynomials of $x$ (see Fig.~\ref{fig:workflow} for the full toolbox). This facilitates quantum model building in the Chebyshev space with polynomials of exponentially large degree, and enables model differentiation. Second, we show how models can be mapped from Chebyshev space to computational basis via a quantum Chebyshev transform (QChT) circuit, enabling the sampling advantage. We apply the developed tools for learning explicit probability distributions, generating samples on extended registers, and discuss its impact on solving differential equations. Our work opens ways for efficient generative modelling, and prompts to reconsider quantum protocols in the Chebyshev space.


\textit{Quantum models.---}We define a quantum model as a circuit-based function that depends on an embedded (explicit) or implicit variable $\bm{x}$ (possibly multidimensional), and parameters $\bm{\theta}$ for an adaptive quantum circuit $\hat{V}_{\bm{\theta}}$. Depending on how variables are treated, there are several types of quantum models, being explicit and implicit models, correspondingly. The explicit models are used in classification \cite{Cerezo2022rev}, regression \cite{mitarai2018quantum}, and scientific machine learning \cite{kyriienko2021solving,Kyriienko2022}. The implicit models are based on measurement results, a prominent example is the quantum circuit Born machine (QCBM), and are routinely used in generative modelling \cite{Liu2018,PerdomoOrtiz2018,Coyle2020}. Below, we define both types of quantum models, and describe how to unify the two concepts with the quantum Chebyshev modelling framework. 

Explicit quantum models rely on a quantum feature map (embedding) \cite{Schuld2019QML,lloyd2020quantum}---a quantum circuit $\hat{\mathcal{U}}_{\varphi}(\bm{x})$ parametrized by $\bm{x}$. Here, $\bm{x}$ denotes a vector of continuous or discrete variables, and in the following we concentrate on single variable case $x$, discussing multivariate generalization towards the end of the letter. Quantum feature maps often take the form of re-uploaded single-qubit rotations \cite{SchuldSweke2021}. This variable-dependent part of the circuit supplies features to the model, which are then adjusted by the adaptive circuit parts $\hat{V}_{\bm{\theta}}$. The explicit quantum model reads $f_{\bm{\theta}}(x) = \mathrm{tr} \{ \hat{\mathcal{U}}_{\bm{\theta},x}^{\dagger} \hat{\rho}_0 \hat{\mathcal{U}}_{\bm{\theta},x} \hat{\mathcal{C}} \}$, where $\hat{\rho}_0$ is an initial state, typically chosen as a density operator corresponding to the zero computational basis state, $\hat{\rho}_0 \coloneqq |\mathrm{\o}\rangle \langle \mathrm{\o}|$, where $|\mathrm{\o}\rangle =  |0\rangle^{\otimes N}$ for $N$ qubits. This is transformed by $\hat{\mathcal{U}}_{\bm{\theta},x}$ as interleaved combination of feature maps and adaptive circuits. The scalar-valued model is then obtained by measuring an observable $\hat{\mathcal{C}}$, a pre-defined cost operator for the circuit read-out. The choice of cost operators in QML remains an open question \cite{Larocca2021overparametrization}, and is typically chosen as simple Z-basis measurement for selected qubits, or projector on certain computational state. We also note that quantum models can involve mid-circuit measurements and trace-out operations, such that adaptive completely positive trace-preserving maps $\mathcal{E}_{\bm{\theta},x} (\hat{\rho}_0)$ are effectively implemented.

Implicit quantum models rely on treating variable $x$ as a measurement result, corresponding to the collapse on a (computational) basis state $|x\rangle$. In this case the model is encoded as a probability distribution $p_{\bm{\theta}}(x) = \mathrm{tr} \{ \hat{V}_{\bm{\theta}}^{\dagger} \hat{\rho}_0 \hat{V}_{\bm{\theta}} |x\rangle \langle x| \}$, following the Born's rule. We note that the embedding disappears, and the overlap with state $\hat{\rho}_0$ transformed by $\hat{V}_{\bm{\theta}}$ solely defines the probability. The distinct part of this model is that each measurement result from the circuit leads to drawing a sample from parameterized probability distribution, $x \sim p_{\bm{\theta}}(x)$. This lays the ground for generative modelling tasks with QCBM-type generators. However, the main difficulty is in learning the ground-truth distribution $p_{\mathrm{truth}}$ such that it is matched by $p_{\bm{\theta}^*}(x)$ at some parameters $\bm{\theta}^*$. Additionally, the overlap models can be exploited implicitly, $g_{\bm{\theta}}(x) = \langle x | \hat{V}_{\bm{\theta}} |\mathrm{\o}\rangle $, which uses SWAP test for real and imaginary parts of the overlap, unlike the absolute value squared case for the probability. 

Recently, the connection between explicit and implicit models was noted~\cite{kasture2022protocols,Kyriienko2022}, and discussed in the context of QML loss functions~\cite{rudolph2023trainability}. Specifically, we observe that for models built on circuits that factorize into sequential feature map and adjustable layers, $\hat{\mathcal{U}}_{\bm{\theta},x} = \hat{V}_{\bm{\theta}} \hat{\mathcal{U}}_{f}(x)$, and projector-type cost operator $\hat{\mathcal{C}} = |\mathrm{\o}\rangle \langle \mathrm{\o}| \eqqcolon \hat{\mathcal{C}}_0$, we can make use of explicitly generated $x$-measurement. For this we demand that the feature map is such that the basis state is embedded as $|x\rangle = \hat{\mathcal{U}}_{f}(x) |\mathrm{\o}\rangle$ for $x \in \mathbb{Z}_{2^N-1}$. Then, the implicit and explicit models match as
\begin{equation}
\label{eq:correspondence}
    p_{\bm{\theta}}(x) = \mathrm{tr} \left\{ \hat{V}_{\bm{\theta}}^{\dagger} \hat{\rho}_0 \hat{V}_{\bm{\theta}} |x\rangle \langle x| \right\} = \mathrm{tr} \left\{ \hat{\mathcal{U}}_{f}^{\dagger}(x) \hat{V}_{\bm{\theta}}^{\dagger} \hat{\rho}_0 \hat{V}_{\bm{\theta}} \hat{\mathcal{U}}_{f}(x) \hat{\mathcal{C}}_{0} \right\} .
\end{equation}
The distribution above can be sampled (left-hand side) or accessed as a scalar valued function at specified variable $x$, after averaging (right-hand side). Importantly, the explicit form of $p_{\bm{\theta}}(x)$ can be differentiated with respect to $x$. This enables including differential constraints \cite{kyriienko2021solving} (used physics-informed machine learning) and studying sensitivities \cite{Stamatopoulos2022towardsquantum} (also known as Greeks in finances). Understanding how to build models in feature (latent) spaces and mapping them to samplable distributions is the key to successful generative modelling.


\textit{Orthonormal Chebyshev map.---}Our goal is designing models based on feature maps that can enable sampling in the computational basis. The computational basis corresponds to the set of orthonormal states $\{ |x_j\rangle \}_{j=0}^{2^N-1}$ such that overlaps equate to the Kronecker delta function, $\langle x_j|x_{j'}\rangle = \delta_{j,j'}$. One can easily generate states $|x_j\rangle$ by applying Pauli $\hat{X}$ operators to the computational zero, sometimes referred to as the basis encoding \cite{schuld2021supervised}. However, building models in this space is difficult, as it assumes adjusting amplitudes (probabilities) for the entire domain. A different approach can be used if the model is built based on another orthonormal set of states, and the appropriate mapping between the two is developed. Given this wish-list, we show how this can be achieved via the orthonormal Chebyshev map and the associated transform.

Specifically, we want to generate an (unnormalized) state $|\tau(x)\rangle$ with amplitudes given by Chebyshev polynomials $T_{k}(x) \equiv \cos(k \arccos(x))$ of the first kind and degree $k$,
\begin{align}
\label{eq:tau(x)}
    |\tau(x)\rangle = \frac{1}{2^{N/2}}T_0(x)|\mathrm{\o}\rangle + \frac{1}{2^{(N-1)/2}} \sum_{k=1}^{2^N-1} T_k(x)|k\rangle,
\end{align}
where we note the distinct amplitude for the computational zero state weighted by $T_0(x) \equiv 1$. Given the properties of polynomials $T_k(x)$ and their orthogonality conditions, 
\begin{align}
\label{eq:T_properties}
    \sum_{j=0}^{2^N-1} T_k(x_j^{\mathrm{Ch}}) T_{\ell}(x_j^{\mathrm{Ch}}) = 
    \begin{cases} 0, ~ &k \neq \ell, \\
    2^N ~ &k = \ell =0, \\
    2^{N-1}~ &k = \ell \neq 0,
    \end{cases}
\end{align}
the states in Eq.~\eqref{eq:tau(x)} are orthonormal at the Chebyshev nodes $x_{j}^{\mathrm{Ch}} \coloneqq \cos\left(\pi(2j+1)/2^{N+1} \right)$, defined at zeros of Chebyshev polynomials. 
Namely, the set of Chebyshev states $\{|\tau(x_{j}^{\mathrm{Ch}})\rangle\}_{j=0}^{2^N-1}$ are such that $\langle \tau(x_{j}^{\mathrm{Ch}})|\tau(x_{j'}^{\mathrm{Ch}})\rangle = \delta_{j,j'}$. We note that the Chebyshev nodes $x_{j}^{\mathrm{Ch}} \in (-1,1)$ form a non-equidistant grid, unlike the standard computational basis and Fourier basis associated to equidistant ``ruler''-type and ``clock''-type grids (see Fig.~\ref{fig:workflow}(a) for the illustration). We also highlight that outside of the Chebyshev nodes the states $|\tau(x)\rangle$ are not orthogonal. Their squared overlap is plotted in Fig.~\ref{fig:workflow}(b), and can be derived analytically for one of the variables $x'$ set to one of the Chebyshev nodes, such that
\begin{equation}
\label{eq:overlap}
    |\langle \tau(x')|\tau(x) \rangle|^2 = \frac{\Big(T_{2^N +1}(x') T_{2^N}(x) - T_{2^N}(x') T_{2^N +1}(x)\Big)^2}{2^{2N} (x' - x)^2},
\end{equation}
which can be derived from the Christofel-Darboux formula for Chebyshev polynomials \cite{Farikhin2011}. Thus, preparing Chebyshev states for any $x \in (-1,1)$ by a unitary quantum circuit requires an ancillary qubit, and leads to normalized states by definition. We denote such normalized states as $|\tilde{\tau}(x)\rangle \coloneqq  |\tau(x)\rangle/ \sqrt{\langle \tau(x)|\tau(x) \rangle}$, which coincide with $|\tau(x)\rangle$ for the Chebyshev nodes, and converge in the large $N$ limit.

To construct the orthonormal Chebyshev feature map $\hat{\mathcal{U}}_{\tau}(x)$ as a circuit which prepares $|\tilde{\tau}(x)\rangle$ state, we observe that Chebyshev polynomials are represented by cosines evaluated at the specific grid. Thus, they can be prepared using a combination of exponents for some scaled variable $x$, $\cos(x) = \{\exp(ix) + \exp(-ix)\}/2$, where each amplitude is embedded via the phase feature map \cite{Kyriienko2022}. The two can be combined using the linear combination of unitaries (LCU) approach \cite{Childs2017}, where maps are conditioned on the state of the (top) ancilla qubit, and effectively interferred, choosing the ancilla collapsed to $0$ outcome. However, since $\exp(-ix) = \exp(ix) \exp(-i2x)$, we can condition only one of the phase feature maps [see corresponding blocks in Fig.~\ref{fig:workflow}(c) labeled as $\exp(\pm ix)$ map]. With this, equal-weight combinations of exponents are prepared as amplitudes for variables scaled such that $T_k(x)$ are recovered. Finally, we need to adjust the weight of the constant term $T_0(x)$. For this we use a round of Grover iterate circuit \cite{NielsenChuang}, rotating the state around $|\mathrm{\o}\rangle$ for the fixed angle. As the measurement operation on the ancilla commutes with rotation, we push it to the end and conclude the circuit.


We highlight that Chebyshev-based feature maps introduced in previous studies \cite{kyriienko2021solving,Paine2021} are fundamentally different, as they are not orthonormal. This prevents their conversion to amplitude encoding (standard in quantum information processing), and building generative models in particular.


\textit{Chebyshev transform.---}Next, we need to develop a map between Chebyshev states and the computational basis states (and reverse). Note that once we have prepared the states forming an orthonormal basis, there exists a bijection and corresponding transformation between this basis and any different orthonormal basis. Here, we introduce the Chebyshev transform as $\hat{\mathcal{U}}_{\mathrm{QChT}} = \sum_{j=0}^{2^N-1}|{\tau(x_{j}^{\mathrm{Ch}})}\rangle \langle x_j|$.
We show the corresponding circuit in Fig.~\ref{fig:workflow}(d), and explain the reasoning below.

First, we note that the Chebyshev transform can be understood as a specific version of the cosine transform \cite{Klappenecker2001}. Namely, the vector of amplitudes for state $|\tau(x_{j}^{\mathrm{Ch}})\rangle$ corresponds to the $(j+1)^{\text{th}}$ column of the type-II discrete cosine transform matrix, DCT$_N^{\mathrm{II}}$, defined as 
\begin{align}
\label{eq:DCT}
\mathrm{DCT}_{N}^{\mathrm{II}} & \coloneqq 2^{-\frac{N-1}{2}}
\Big\{c_k \cos[k(j+1/2)\pi]/2^N \Big\}_{k,j=0...2^N-1},
\end{align}
where $c_0 = 1/\sqrt{2}$, and $c_k = 1$ for $k \neq 0$. We note that this matrix is strongly related to Fourier transform matrix, but requires interfering and mixing its elements, thus suggesting the use of an extended QFT circuit. The circuit starts with a Hadamard gate on the ancilla, being the most significant bit, followed by a CNOT ladder that is typically used for a cat state preparation. This is followed by a $N+1$ QFT circuit, which is later converted into blocks of purely real and imaginary components through a series of unitary gates. Namely, the single qubit gates $\mathrm{U}_1$ and $\mathrm{R}_\mathrm{Z}$ are introduced to adjust the relative phases for states split as $|0\rangle_a|\Phi\rangle$  and $|1\rangle_a|\Phi\rangle$, for any $N$-qubit intermediate state denoted by $|\Phi\rangle$. The permutation circuit is used to reorder amplitudes of the conditioned states, 
followed by another CNOT ladder. The circuit is concluded with the constant phase $\mathrm{U}_2$ and multi-controlled $\mathrm{R}_\mathrm{X}$ gates to fix weights and to ensure that the amplitudes of $|0\rangle_a$$|\Phi\rangle$ ($|1\rangle_a$$|\Phi\rangle$) are purely real (imaginary) for any input states.  
We note that the ancilla starts and ends in $|0\rangle$ state (``clean'' run). Finally, concatenating the described embedding and the map, we get $\hat{\mathcal{U}}_{f}(x) = \hat{\mathcal{U}}_{\mathrm{QChT}}^\dagger \hat{\mathcal{U}}_{\tau}(x)$.


\textit{Chebyshev map derivative.---}Next, we briefly discuss how to differentiate Chebyshev models. 
This can be approached in several ways. First, we can use the parameter shift rule~\cite{schuld2019evaluating,mitarai2018quantum,kyriienko2021generalized}, which is valid in the training stage when cost function corresponds to projection on zero state (including ancilla qubit). In this case we need to decompose controlled phase rotations using CNOT conjugation, and apply two shifts per individually reuploaded parameter $x$. 
In total, the orthonormal Chebyshev map can be differentiated with $4N$ shifts. Alternatively, one can see this as differentiation in presence of mid-circuit measurements~\cite{wiersema2021measurementinduced}. Another option is to differentiate the feature map formally, extracting an effective generator $\hat{\mathcal{G}}_{\mathrm{eff}}$, and taking derivatives as a linear combination of unitaries~\cite{schuld2019evaluating,Mitarai2019Htest,Paine2023} (see Supplementary Note). 
\begin{figure}[t]
\begin{center}
\includegraphics[width=1.0\linewidth]{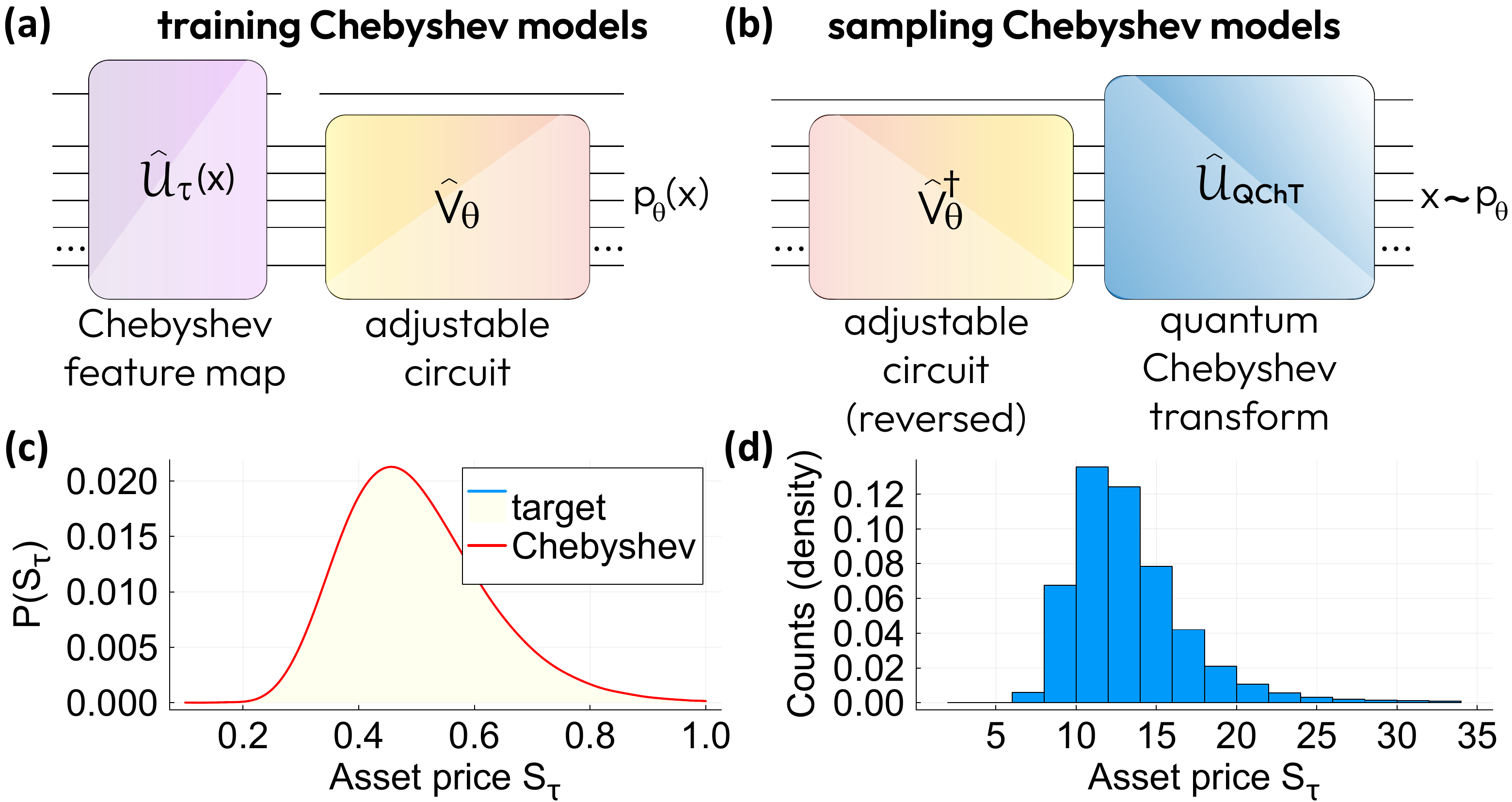}
\end{center}
    \caption{\textbf{Learning and sampling distributions with quantum Chebyshev models. (a)} Circuit used to train the model in latent space. \textbf{(b)} Circuit used to sample the model in the computational basis. \textbf{(c)} Training a lognormal distribution with parameters $\mu=0$, $\sigma=0.25$, $S_0=0.5$, and $t_0=1$. \textbf{(d)} Sampling from the trained lognormal distribution, generated with $10^6$ shots (binned).}
\label{fig:results}
\end{figure}


\textit{Learning and sampling: results.---}Next, we demonstrate examples of applying quantum Chebyshev transform to generative modelling from relevant distributions. We show how to learn and sample a distribution of underlying asset prices described by the Black-Scholes pricing model \cite{black_scholes}. We consider an asset whose price follows a geometric Brownian motion under the risk-neutral measure. At time $t$, the price $S_t$ (stochastic variable) is governed by the stochastic differential equation $dS_t=\mu S_tdt+\sigma S_tdW_t$ with constant drift $\mu$, constant volatility $\sigma$ and stochastic process $W_t$. Applying the Black-Scholes model gives the solution $\ln(S_t)=\ln{S_0}+(\mu-\sigma^2/2)t+\sigma W_t$, implying that the underlying asset price is distributed lognormally with probability density function
\begin{equation}
    \mathbb{P}(S_t) = \frac{1}{S_t\sigma\sqrt{2\pi t}}\text{exp} \left\{ \frac{-[\text{ln}(S_t/S_0)+(\mu-\sigma^2/2)t]^2}{2\sigma^2t} \right\}.
\end{equation}
Setting the $\mathbb{P}(S_{t \rightarrow t_0}) = p_{\mathrm{truth}}$ as a ground truth, we use the Chebyshev model to learn this distribution. 
To do this, we utilise the differentiable quantum generative models (DQGM) framework \cite{Kyriienko2022}. We begin by variationally training the lognormal distribution in the latent space with the Chebyshev feature map $\hat{\mathcal{U}}_\tau(x)$ shown in Fig.~\ref{fig:results}(a), and then sample from the distribution in the computational basis using the Chebyshev transform circuit $\hat{\mathcal{U}}_\text{QChT}$ shown in Fig.~\ref{fig:results}(b). The result in Fig.~\ref{fig:results}(c) shows the trained lognormal distribution using $N=5$ qubits with a variational hardware efficient ansatz $\hat{V_\theta}$ of depth $14$ \cite{Paine2021}. We employ a mean squared error loss with a low learning rate of $0.005$, simulated over five thousand epochs using Julia's Yao package \cite{YaoFramework2019}. The training grid consists of positive Chebyshev nodes $\{ x_j^{\text{Ch}} \}~\forall \text{ }j\in[0,2^{N-1}]$ and additional points between these nodes $\{ x_{j/2}^{\text{Ch }}\}$. The histogram in Fig.~\ref{fig:results}(d) shows the resulting sampled distribution.
The success of sampling is defined by several properties of the models used during learning. This includes expressivity (ability of basis set and tunable coefficients to represent a distribution) and trainability of the variational circuit. The interplay between the two largely defines distributions that can be learnt.

\begin{figure}[t]
\begin{center}
\includegraphics[width=1.0\linewidth]{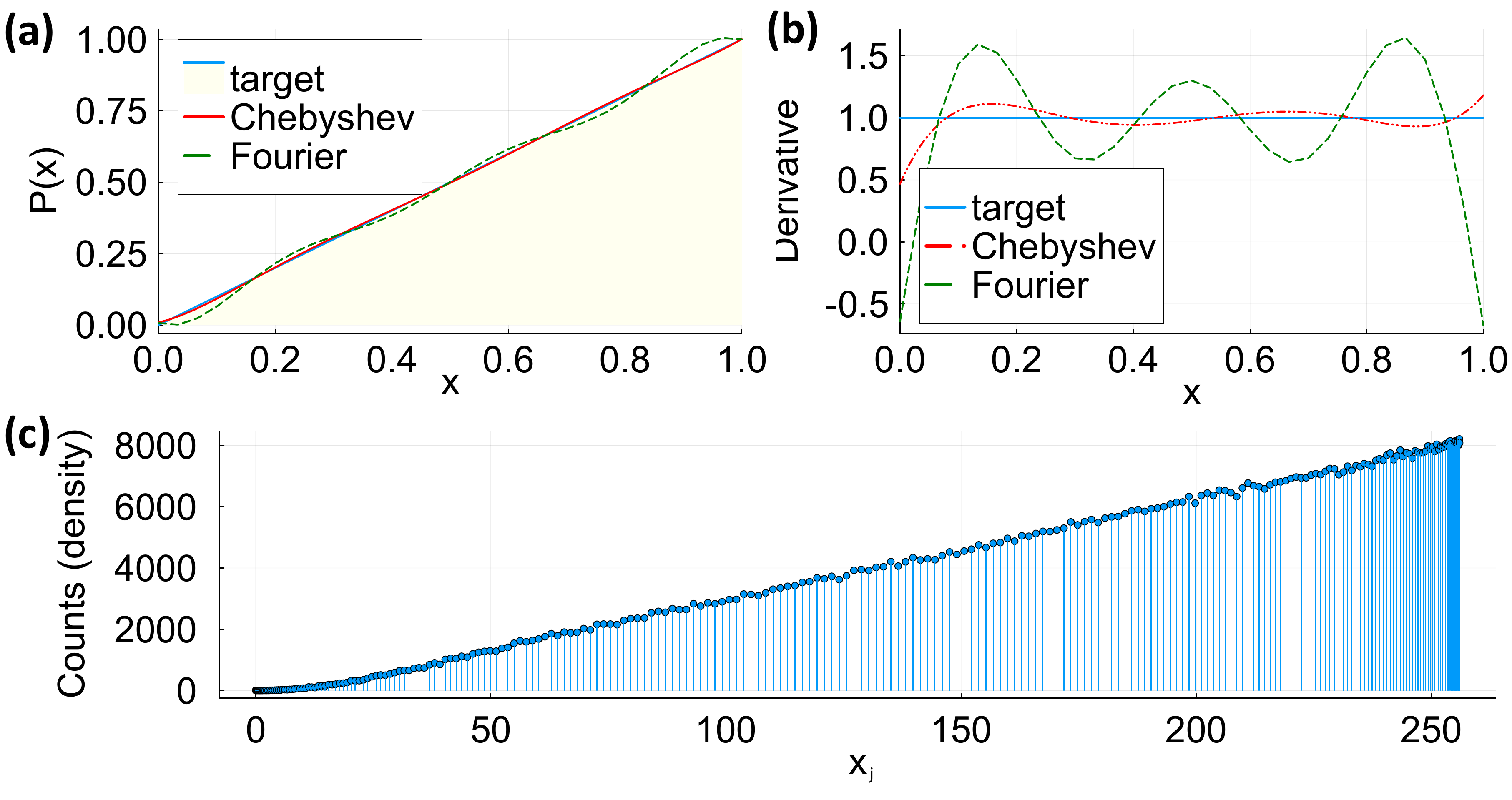}
\end{center}
    \caption{\textbf{Comparison between the quantum Chebyshev model and the quantum Fourier model for generative modelling using the DQGM framework. (a)} Training a linear distribution with ansatz depth of 6 layers. \textbf{(b)} Derivatives of the trained linear models. \textbf{(c)} Sampling from the trained linear distribution on an extended register of 8 qubits, shown for integer labels of Chebyshev nodes $x_j$.}
\label{fig:results_linear}
\end{figure}

As a further example, we consider a linear distribution with probability density function $\mathbb{P}(x)=x$. In this case, we highlight the importance of the Chebyshev basis as compared to the Fourier basis of similar expressivity. Using only $N=2$ qubits for the embedding, we learn the linear distribution with the orthonormal Chebyshev, and compare it to the Fourier model built with the phase map \cite{Kyriienko2022}. The results are shown in Fig. \ref{fig:results_linear}(a), where hyperparameters are same as before. We note that while Chebyshev model follows $\mathbb{P}(x)$ closely, the Fourier model experiences oscillations around the linear trend. Moreover, once we evaluate the derivatives for the respective generative models, we observe that while Chebyshev's derivative comes close to one, the Fourier derivatives are largely off [Fig. \ref{fig:results_linear}(b)]. This is an important consideration for solving differential equations, as large deviation in derivative-based loss terms lead to poor convergence overall. The plot in Fig. \ref{fig:results_linear}(c) shows the sampled distribution of the linear quantum Chebyshev model, where we use the optimized $\hat{V}_{\bm{\theta}^*}$ for 2 qubits, and map it to an extended register of $N=8$ qubits with QChT. We remind that samples are shown for the Chebyshev grid.

\textit{Discussion.---}The proposed framework for building quantum models based on Chebyshev models can be used in various ways. First, as shown above, due to orthonormality and mapping to computational basis the prepared models are readily sampled and extended over larger registers \cite{Garcia-Molina2022,Kyriienko2022}. Second, the preparation of the implicit model can be seen as loading of the amplitude encoded state \cite{Rattew2021efficient,rattew2022preparing}, which is required for solving systems of linear (and differential) equations \cite{harrow2009quantum,Subasi2019}. Third, the toolkit can be used for solving differential equations, both in the latent space and computational space, making the bridge between the two. Specifically, the Chebyshev basis is highly preferred for classical spectral methods \cite{boyd2001}, and is \emph{de facto} a standard in approximation theory \cite{trefethen2019approximation}. Also, the mapping allows simple incorporation of the boundary conditions with explicit encoding --- the problem that has been noted when dealing exclusively with amplitude encoding \cite{Childs2020} (i.e. implicit models). Fourth, quantum Chebyshev models can be readily employed in explicit QML models concerning classification \cite{Schuld2019QML,Chen2021}, regression \cite{mitarai2018quantum,Paine2023}, and anomaly detection \cite{kyriienko2022unsupervised,Grossi2022}, both in the variational and kernel-based settings (and targeting cases where high model expressivity is required). 
Other extensions to the described framework correspond to multi-dimensional modelling. For this we can employ parallel reuploading, where $\bigotimes_{\ell=1}^{L} \hat{\mathcal{U}}_{\tau}(x_\ell)$ embeds dependence on the vector of variables $\bm{x} = (x_1, ..., x_L)$. The maps can follow the tensorial structure, and models can be either via shared preparation unitary $\hat{\bm{V}}_{\bm{\theta}}$ acting on $L N$ qubits, or exploit copula-type models when dealing with multivariate probability distributions \cite{Zhu2022,Kyriienko2022}. 

Finally, we can connect the quantum model building with conventional protocols outside of QML remit. We note that Hamiltonian dynamics $\exp(-i t \mathcal{\hat{H}})$ can also be seen as an embedding of the time variable $t$, with features defined by the spectrum of the Hamiltonian $\mathcal{\hat{H}}$. In this case we get an explicit quantum model with spectral features in the Fourier space. This is usually processed via QFT for reading out relevant energies. However, alternative quantum walk operators in the modified form of $\exp(i \arccos[t \mathcal{\hat{H}}])$, leading to significant computational improvements~\cite{BabbushPRX2018}. We note that from QML side, this quantum walk operators implicitly embed Chebyshev-type features. Transforming them into bit-basis states is another motivation of developing quantum Chebyshev transform, and we consider studying this relation as an important avenue of future research.


\textit{Conclusions.---}We developed a protocol for building quantum models based on Chebyshev polynomials $T_k(x)$ being the quantum state amplitudes. Similar to quantum Fourier transform that maps computational basis space into the phase basis, we described the quantum circuit for mapping between computational and Chebyshev basis spaces. We proposed an $x$-dependent embedding circuit for generating the exponentially expressive orthonormal Chebyshev basis. This enables the quantum model differentiation. Among various possible applications, we applied quantum Chebyshev transform to generative modeling, for distributions being solutions of stochastic differential equations. Learning physically- and financially-motivated distributions, we performed quantum Chebyshev transform for efficient sampling of these distributions in extended computational basis.\vspace{1mm}

%

\clearpage
\newpage 
\appendix

\section{SUPPLEMENTARY NOTE}

\textbf{Evaluating derivatives: LCU approach.---}Here, we expand on how to calculate the derivatives of the Chebyshev encoding. One such way is with use of an effective operator $\hat{\mathcal{G}}_{\mathrm{eff}}$. For this we consider the (not necessarily normalized) state $|\tau(x) \rangle$ from Eq.~\eqref{eq:tau(x)} in the main text, where all the amplitudes are scaled Chebyshev polynomials. The normalised version of this $|\tilde{\tau}(x)\rangle$ is prepared by the Chebyshev feature map. When taking the derivative with respect to $x$ of model containing $|\tau(x)\rangle$, the resulting model contains a derivate state which we done as $|\tau'(x)\rangle$. We can then consider this a state with each amplitude as scaled Chebyshev polynomial derivatives.

We note that the Chebyshev polynomials and their derivatives are related. It is known that $T_n'(x) = n U_{n-1}(x)$, where $U_n(x)$ are Chebyshev polynomials of the second kind \cite{wang2015some}. This can then be further expanded to
\begin{align}
    T_0'(x) &= 0, \\
    T_{2n}'(x) &= 4n \sum_{m=1}^n T_{2m-1}(x), \\
    T_{2n+1}'(x) &= (4n+2) \sum_{m=1}^n T_{2m}(x) + (2n+1)T_0(x).   
\end{align}
Therefore the derivative of a Chebyshev polynomial is a linear combination of lower order Chebyshev polynomials. From this we can postulate there exists an effective operator $\hat{\mathcal{G}}_{\mathrm{eff}}$ such that
\begin{align}
    |\tau'(x)\rangle = \hat{\mathcal{G}}_{\mathrm{eff}}  |\tau(x)\rangle.
\end{align}
We note that $\hat{\mathcal{G}}_{\mathrm{eff}}$ is non-unitary and shall be implemented by techniques suitable for embedding non-unitary operators (such as linear combination of unitaries) and exploiting an extended Hilbert space.

We now explore how this is used for the Chebyshev model as used in the results section. We have
\begin{align}
    p_\theta(x) &= |\langle \tau(x) | \psi_\theta \rangle|^2, \\ \notag
                &= \mathcal{N}(x)\mathcal{N}^\dagger(x)|\langle \tilde{\tau}(x) | \psi_\theta \rangle |^2.
\end{align}
Here $|\tilde{\tau}(x)\rangle = |\tau(x)\rangle/\sqrt{\langle \tau(x)|\tau(x)\rangle}$, and we introduce the notation $\mathcal{N}(x) = \sqrt{\langle \tau(x)|\tau(x)\rangle}$. $\mathcal{N}(x)$ can thus be evaluated as $\mathcal{N}(x) = \left(1/2 + \sum_{j=1}^{2^N-1} T_j(x)^2 \right)^{1/2}/2^{(N-1)/2}$. The derivative (using the product rule) is then
\begin{align}
    dp_\theta(x)/dx = ~ &\langle \tau'(x) | \psi_\theta \rangle \langle \psi_\theta| \tau(x) \rangle + \langle \tau(x) | \psi_\theta \rangle \langle \psi_\theta| \tau'(x) \rangle, \\ \notag
    = ~&\langle \tau(x) | \hat{\mathcal{G}}^\dagger_{\mathrm{eff}} | \psi_\theta \rangle \langle \psi_\theta| \tau(x) \rangle  \\ \notag
    &+ \langle \tau(x) | \psi_\theta \rangle \langle \psi_\theta| \hat{\mathcal{G}}_{\mathrm{eff}} | \tau(x) \rangle, \\ \notag
    = ~&\mathcal{N}(x)\mathcal{N}^\dagger(x) \big(\langle \tau(x) | \hat{\mathcal{G}}^\dagger_{\mathrm{eff}} | \psi_\theta \rangle \langle \psi_\theta| \tau(x) \rangle \\ \notag
    &+ \langle \tau(x) | \psi_\theta \rangle \langle \psi_\theta| \hat{\mathcal{G}}_{\mathrm{eff}} | \tau(x) \rangle\big). \notag
\end{align}
Since any operator can be expanded into a sum of unitaries, the derivative of Chebyshev model can be evaluated by calculating weighted overlaps specified above.

\end{document}